%% file: paper.tex
\documentclass[usenatbib]{mnras}
\usepackage{upgreek} %
\usepackage{amsmath} %
\usepackage{amssymb} %
\usepackage{bm} %
\usepackage{hyperref} %
\usepackage[english]{babel} %
\usepackage{listings} %
\usepackage{IEEEtrantools} %
\usepackage[pdftex]{graphicx} %

\input{journals.tex}

\newcommand{\expo}[1]{\mathrm{e}^{#1}} 
\newcommand{\ud}{\mathrm{d}} 

\newcommand{\eqn}[1]{\text{equation (\ref{#1})}}
\newcommand{\fig}[1]{\text{Figure \ref{#1}}}
\newcommand{\sect}[1]{\text{\S\ref{#1}}}

\newcommand{\mtd}{$<$3D$>$} 
\newcommand{\cf}{\mathcal{C}_{\nu}}
\newcommand{\cfpp}{\mathcal{C}^{\text{pp}}_{\nu}}
\newcommand{\cfint}{\mathcal{C}}

\newcommand{\arga}{\left(\bm{r};\Omega\right)}
\newcommand{\argument}{\left(\bm{r};\Omega\right)}

\newcommand{\fluxdep}{\mathcal{A}_{\nu}}
\newcommand{\linestrength}{\mathcal{A}}

\newcommand{\intdep}{D_{\nu}}
\newcommand{\intdeppert}{D^{1}_{\nu}}

\newcommand{\intcont}{I^{\mathrm{c}}_{\nu}}
\newcommand{\inttot}{I_{\nu}}
\newcommand{\intbroad}{I^{\text{broad}}_{\nu}}

\newcommand{\broad}[1]{\mathcal{B}\left[#1\right]}
\newcommand{\broadarg}{\mathcal{B}}

\newcommand{\exttot}{\alpha_{\nu}}
\newcommand{\extcont}{\alpha^{\mathrm{c}}_{\nu}}
\newcommand{\extline}{\alpha^{\mathrm{l}}_{\nu}}

\newcommand{\sftot}{S_{\nu}}
\newcommand{\sfcont}{S^{\mathrm{c}}_{\nu}}
\newcommand{\sfline}{S^{\mathrm{l}}_{\nu}}
\newcommand{\sfdep}{S^{\mathrm{eff}}_{\nu}}
\newcommand{\sfdeppert}{S^{\mathrm{eff},1}_{\nu}}

\newcommand{\opdepth}{\tau_{\nu}}
\newcommand{\opdepthr}{\tau^{\text{r}}_{\nu}}
\newcommand{\lgtfive}{\log_{10}\tau^{\text{r}}_{500}}

\newcommand{\triplet}{\text{OI\,777\,nm}}

\newcommand{\rf}{\mathcal{R_{\nu}}}
\newcommand{\rfint}{\mathcal{R}}

\newcommand{\mathpi}{\uppi}

\title[3D line contribution functions]{On line contribution functions 
and examining spectral line formation in 3D model stellar atmospheres}
\author[A.~M.~Amarsi]{A.~M.~Amarsi$^{1}$\thanks{E-mail: anish.amarsi@anu.edu.au}\\
$^{1}$Mount Stromlo Observatory, Australian National University,
Weston Creek, ACT 2611, Australia}
\date{Accepted 2015 June 22. Received 2015 June 13; in original form 2015 February 16}
\pagerange{\pageref{firstpage}--\pageref{lastpage}} \pubyear{---}
\begin{document}
\maketitle 
\label{firstpage}
\begin{abstract}
Line contribution functions are useful
diagnostics for studying spectral line formation
in stellar atmospheres.
I derive an expression for the contribution function
to the absolute flux depression
that emerges from three-dimensional
`box-in-a-star' model stellar atmospheres.
I illustrate the result 
by comparing the local thermodynamic equilibrium (LTE)
spectral line formation of the high-excitation permitted \triplet~lines
with the non-LTE case.

\end{abstract}
\begin{keywords}
line: formation -- radiative transfer -- methods: numerical -- stars: atmospheres 
\end{keywords}
\section{Introduction}
\label{introduction}

The fundamental parameters of stars such as
their effective temperature,
surface gravity, and chemical composition
are not observable quantities:
rather, they must be inferred using model
stellar atmospheres
\citep{Bergemann:2014aa}.
Three dimensional (3D) hydrodynamic 
`box-in-a-star' models 
\citep{1982A&amp;A...107....1N}
are increasingly being used in this context
\citep{2009MmSAI..80..711L,2013A&amp;A...557A..26M,2013ApJ...769...18T}.
These present a huge improvement 
over classical 1D hydrostatic models
on account of their ab initio treatment
of convective energy transport in the outer envelope
that can realistically reproduce
the shifting, broadening and strengthening
of spectral lines by convective velocity fields
and atmospheric inhomogeneities 
 \citep{1980LNP...114..213N,1999A&amp;A...346L..17A}.
Inferred logarithmic abundances 
can suffer errors as large as
 $\pm1.0\,\mathrm{dex}$ when modelled in 1D
\citep{2008MmSAI..79..649C}.

Visualizing and
understanding spectral line formation 
in three dimensions is non-trivial.
Contribution functions
\citep{1952hss..book.....D,1974SoPh...37...43G}
are useful tools to that end.
They can be interpreted 
as probability density functions for 
line formation in the atmosphere
\citep{1972SoPh...24..255S,1986A&amp;A...163..135M}
and are often used to infer the mean formation depths 
of spectral lines.
The line intensity contribution function 
\citep{1986A&amp;A...163..135M}
represents the contribution from different
locations of the atmosphere to the 
depression in the normalized intensity.
This quantity is commonly used to study lines 
in a solar context
\citep{2008A&amp;A...488.1031C}.
Since stars are in general not resolved, 
often more relevant is
the line flux contribution function, 
\citep{1996MNRAS.278..337A},
which is instead formulated in terms of 
the depression in the absolute flux. 

Since all parts of the stellar atmosphere contribute
to its observed flux profile,
the line flux contribution function
is a function of 3D space.
\citet{1996MNRAS.278..337A} derive
it in the context of 1D model stellar atmospheres,
i.e. assuming plane-parallel symmetry.
To apply it directly to a 3D model would 
be to treat the atmosphere
as an ensemble of 1D columns 
i.e.~it would be a 1.5D approximation
\citep{1995A&amp;A...302..578K}.
This is undesirable because 
the effects of horizontal radiative transfer
are entirely neglected.
Another approach is
to compute the plane-parallel contribution function
on a horizontally-averaged, \mtd~model.
This approach is still not ideal,
because it neglects the effects of the
atmospheric inhomogeneities 
which characterize real stellar atmospheres.

In this paper I present
in \sect{method} a derivation for 
the line flux contribution function 
that is valid in three dimensions.
To illustrate the result,
I explore in \sect{example} the formation of 
the high excitation permitted \triplet~lines
in a 3D hydrodynamic \textsc{stagger} 
model atmosphere \citep{2013A&amp;A...557A..26M}.
I present a short summary in \sect{conclusion}.

\section{The 3D line flux contribution function} 
\label{method}

\begin{figure}
\begin{center}
\includegraphics[scale=0.26]{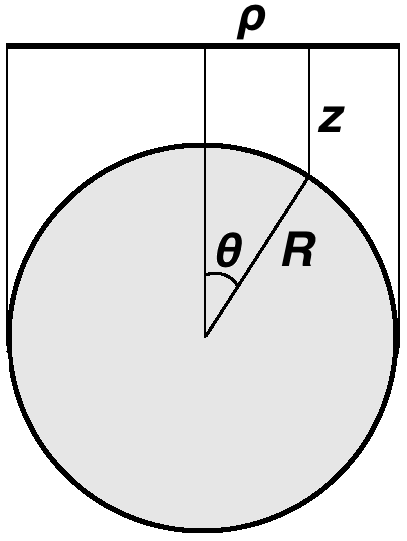}\,\,\,\,\,\,\,\,\,\includegraphics[scale=0.26]{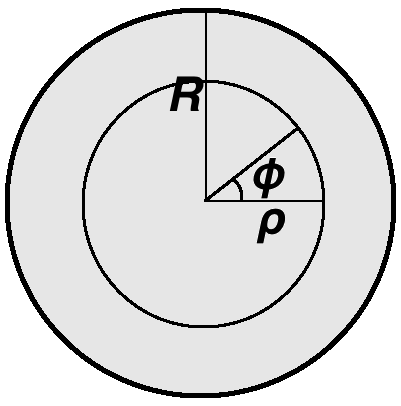}
\caption{Visual aids to the derivation 
presented in \sect{method}.
Left: spherical star and an arbitrary reference plane
perpendicular to the observer's line of sight;
the observer is located to the top of the diagram.
The spherical polar angle is $\theta$;
conventionally, the notation $\mu=\cos\theta$
is adopted.
The cylindrical polar radius is $\rho=R\sin\theta$,
where $R$ is the radius of the star. 
$z$ is the displacement from a point in the atmosphere
to the arbitrary reference plane.
Right: spherical star as seen by the observer,
who is now located out of the page. 
The azimuthal angle is $\phi$.
A ring of constant $\rho$ is shown.}
\label{fig1}
\end{center}
\end{figure}

\subsection{Concept}
\label{concept}

The flux depression at frequency $\nu$
from a star of radius $R$ measured by a distant observer
is proportional to the total emergent intensity depression,
\phantomsection\begin{IEEEeqnarray}{rCl}
\label{eq1}
	\fluxdep &\propto& \frac{1}{\mathpi R^{2}}\int \int \left(\intcont-\inttot\right) \, \rho\,\ud\rho\,\ud\phi\, ,
\end{IEEEeqnarray}
in the cylindrical polar coordinate system 
depicted in \fig{fig1}:
the polar axis intersects disc-centre
and is directed towards the observer.
$\inttot$ is the specific intensity
and $\intcont$ is the specific continuum intensity,
at some position on the disc, 
in the direction of the observer.
The line flux contribution function $\cf$ must satisfy
\phantomsection\begin{IEEEeqnarray}{rCl}
\label{eq2}
    \fluxdep &=& \int_{\text{box}} \cf\left(\bm{r}\right)\,\ud^{3}r\, .
\end{IEEEeqnarray}
Crucially, the integration is not performed over
vertical height as in the 
plane-parallel derivation of \citet{1996MNRAS.278..337A},
but over the entire 3D volume in which the line may form.
This is the entire volume of the 3D model atmosphere;
$\bm{r}$ thus specifies a position in this box.

In what follows, \eqn{eq1} is manipulated 
into the form of \eqn{eq2}, and thereby $\cf$ is inferred.
The contribution function $\cfint$ to the
integrated line strength $\linestrength=\int\fluxdep\,\ud\nu$
is also found, and satisfies 
$\cfint=\int\cf\,\ud\nu$.

\subsection{Derivation}
\label{derivation}

Along any given ray,
$\inttot$ and $\intcont$
satisfy the respective transport equations,
\phantomsection\begin{IEEEeqnarray}{rCl}
\label{eq3}
	\frac{\ud \inttot}{\ud z} &=& \exttot\,\left(\sftot-\inttot\right)\, , \\
\label{eq4}
	\frac{\ud \intcont}{\ud z} &=& \extcont\,\left(\sfcont-\intcont\right)\, , 
\end{IEEEeqnarray}
where $z$ is the path distance,
increasing upward towards the observer.
The linear extinction coefficient $\exttot$
and the source function $\sftot$ are, 
in terms of their line and continuum components,
\phantomsection\begin{IEEEeqnarray}{rCl}
\label{eq5}
    \exttot &=& \extcont+\extline\, ,\\
\label{eq6}
	\sftot &=& \frac{\extcont\,\sfcont + \extline\,\sfline}{\exttot}\, 
\end{IEEEeqnarray}
\citep{2014tsa..book.....H}.

Following \citet{1986A&amp;A...163..135M},
an effective transport equation
for the intensity depression $\intdep\equiv\intcont-\inttot$
is found by subtracting \eqn{eq3} from \eqn{eq4},
\phantomsection\begin{IEEEeqnarray}{rCl}
\label{eq7}
	\frac{\ud \intdep}{\ud z} &=& \exttot\,\left(\sfdep-\intdep\right) \, ,
\end{IEEEeqnarray}
where the effective source function is
\phantomsection\begin{IEEEeqnarray}{rCl}
\label{eq8}
    \sfdep &=& \frac{\extline}{ \exttot } \left(\intcont-\sfline\right)\, .
\end{IEEEeqnarray}
In terms of the optical depth along the ray
$\ud\opdepth=-\exttot\,\ud z\,$,
\eqn{eq7} is expressed as
\phantomsection\begin{IEEEeqnarray}{rCl}
\label{eq9}
	\frac{\ud\intdep}{\ud \opdepth} &=& \intdep-\sfdep\, .
\end{IEEEeqnarray}
The formal solution is found 
by integrating from $\opdepth=0$ to $\opdepth\rightarrow\infty$,
\phantomsection\begin{IEEEeqnarray}{rCl}
\label{eq10}
	\intdep &=& \int \sfdep \, \expo{-\opdepth}\, \ud\opdepth \, .
\end{IEEEeqnarray}
Neglecting proportionality factors,
the flux depression is obtained by
substituting \eqn{eq10} into \eqn{eq1},
\phantomsection\begin{IEEEeqnarray}{rCl}
\label{eq11}
	\fluxdep &=& \int \int \int \exttot\, \sfdep\, \expo{-\opdepth}\, \ud z\,\rho\,\ud\rho\,\ud\phi\, ,
\end{IEEEeqnarray}
where the integrand is evaluated 
with the constraint that the emergent rays are directed 
towards the observer.
As the observer is very far from the star,
the emergent rays are parallel to each other. 
Consequently, the last equation is written
in terms of an infinitesimal volume element,
\phantomsection\begin{IEEEeqnarray}{rCl}
\label{eq12}
	\fluxdep &=& \int_{\text{star}} \exttot\, \sfdep\, \expo{-\opdepth}\, \ud^{3}r\, .
\end{IEEEeqnarray}

The integration in \eqn{eq12} is performed over 
the entire volume of the star.
3D box-in-a-star models of stellar atmospheres
have Cartesian geometry
and span a minute surface area of the
stars they represent
\citep{2012JCoPh.231..919F,2013A&amp;A...557A..26M}.
The flux spectrum
from the modelled star is
(approximately) reproduced by shifting the box
tangentially across the spherical surface. 
This is represented by
two integrations: one over the volume of the box
and the other over the unit hemisphere.
Again neglecting proportionality factors,
\phantomsection\begin{IEEEeqnarray}{rCl}
\label{eq13}
	\fluxdep &\approx& \int \int_{\text{box}} \exttot\arga\, \sfdep\arga\, \expo{-\opdepth\arga}\, \ud^{3}r\, \ud\Omega \, ,
\end{IEEEeqnarray}
where the functional dependence
of the integrand has been made explicit
for clarity. 
The position vector $\bm{r}$ specifies a position within the box,
and the solid angle $\Omega$
specifies the 
direction of the emergent rays.
The infinitesimal solid angle satisfies
$\ud\Omega=\ud\mu\,\ud\phi$,
where $\mu=\cos\theta$.
After changing the order of integration,
the contribution function is inferred to be,
\phantomsection\begin{IEEEeqnarray}{rCl}
\label{eq14}
	\cf\left(\bm{r}\right) &=& \int \exttot\arga\, \sfdep\arga\, \expo{-\opdepth\arga}\, \ud\Omega\, .
\end{IEEEeqnarray}
This represents the contribution of
\emph{a point within the box} to the 
observed absolute flux depression in the line,
at frequency $\nu$. 
The integrated line strength contribution function follows 
immediately,
\phantomsection\begin{IEEEeqnarray}{rCl}
\label{eq15}
	\cfint\left(\bm{r}\right) &=& \int \int \exttot\arga\, \sfdep\arga\, \expo{-\opdepth\arga}\, \ud\Omega\, \ud\nu \, .
\end{IEEEeqnarray}

\subsection{Rotational broadening}
\label{rotational broadening}
Line broadening caused by the rigid rotation of the star 
must be included during post-processing.
This broadening will affect the monochromatic
quantity $\fluxdep$ and hence $\cf$.
Following 
\citet{1990A&amp;A...228..203D},
the broadened specific intensity is,
\phantomsection\begin{IEEEeqnarray}{rCl}
\label{eq16}
	\intbroad &=& \broad{\inttot}\, ,
\end{IEEEeqnarray}
where $\broadarg$ is a functional
which broadens its argument according to,
\phantomsection\begin{IEEEeqnarray}{rCl}
\label{eq17}
	\mathcal{B}\left[x\left(v,\theta,\phi\right)\right] &=& \frac{1}{2\mathpi}\int x\left(v-V\sin\iota\sin\theta\cos\psi,\theta,\phi\right) \, \ud\psi\, .
\end{IEEEeqnarray}
Here $v=c\frac{\Delta\nu}{\nu}$
is the Doppler speed,
$V$ is the rotation speed of the star in the
line forming region,
$\iota$ is the inclination angle of the rotation axis
with respect to the observer, 
and the integral is over an interval of $2\mathpi$.
Retracing the steps above,
one obtains a rotationally-broadened
contribution function,
\phantomsection\begin{IEEEeqnarray}{rCl}
\label{eq18}
    \cf\left(\bm{r}\right) &=& \int \broad{\exttot\arga\, \sfdep\arga\, \expo{-\opdepth\arga}} \, \ud\Omega\, .
\end{IEEEeqnarray}
(In deriving this expression, it is necessary to 
move $\mathcal{B}$ within the integral
of \eqn{eq10}.
This is valid because the atmosphere is 
assumed to be sufficiently shallow that
$V$ does not vary across its depth.)

This integrated line strength $\linestrength$,
should not be affected by the rotation of the star 
\citep{1992oasp.book.....G}.
The adopted broadening formalism is consistent
with this: integrating \eqn{eq17} across the line profile,
\phantomsection\begin{IEEEeqnarray}{rCl}
\label{eq19}
	\int \mathcal{B}\left[x\left(v,\theta,\phi\right)\right] \, \ud v &=& \int x\left(v,\theta,\phi\right)\,\ud v\, ,
\end{IEEEeqnarray}
which implies that the contribution function $\cfint$
is not affected by the rotation of the star.

\subsection{Mean formation depth}
\label{mean formation depth}

The interpretation of the contribution function as a
probability density function for line formation
\citep{1972SoPh...24..255S,1986A&amp;A...163..135M}
suggests a formalism for 
defining the mean formation value
of some quantity $q$ with respect to a line,
\phantomsection\begin{IEEEeqnarray}{rClrCl}
\label{eq20}
    \mathbb{E}\left[q\right] &=& \frac{\int q\left(\bm{r}\right) \,\cfint\left(\bm{r}\right) \, \ud^{3} r}{\int \cfint\left(\bm{r}\right)\, \ud^{3} r}\, ,
\end{IEEEeqnarray}
and the variance might then
be defined in the usual way as $\mathbb{E}\left[q^{2}\right]-\mathbb{E}\left[q\right]^{2}$.
For example, $\mathbb{E}\left[q=\lgtfive\right]$ 
may be used to define the mean formation depth,
where $\lgtfive$ is the logarithmic radial optical depth at 
wavelength $\lambda=500\,\mathrm{nm}$,
a standard measure of depth
in stellar atmospheres.

\subsection{Relationship to the line flux response function}
\label{line response functions}

A related spectral line formation diagnostic is
the response function:
the linear response of the line to a perturbation in the
atmosphere
\citep{1971SoPh...20....3M,1975SoPh...43..289B,1977A&amp;A....54..227C}.
The line flux response function $\rf$ must satisfy
\phantomsection\begin{IEEEeqnarray}{rCl}
\label{eq22}
	\delta \fluxdep &\equiv& \int \rf\left(\bm{r}\right)\,\delta\beta\left(\bm{r}\right) \,\mathrm{d}^{3} r\, ,
\end{IEEEeqnarray}
where $\beta$ is an 
atmospheric parameter (such as temperature). 

Following \citet{1986A&amp;A...163..135M},
the response function is obtained
by adapting the above derivation.
The effective transport equation 
\eqn{eq7} is perturbed 
so that $\intdep\rightarrow \intdep+\delta\beta\,\intdeppert$,
and the equation for $\intdeppert$ is solved,
\phantomsection\begin{IEEEeqnarray}{rCl}
\label{eq23}
	\frac{\ud\intdeppert}{\ud z} &=& \exttot\left(\sfdeppert - \intdeppert\right) ,
\end{IEEEeqnarray}
where the perturbed effective source function is,
\phantomsection\begin{IEEEeqnarray}{rCl}
\label{eq24}
    \sfdeppert &=& \frac{\partial \sfdep}{\partial\beta} + \frac{1}{\exttot}\,\frac{\partial\exttot}{\partial\beta}\, \left(\sfdep-\intdep\right).
\end{IEEEeqnarray}
The response function is then found
by following the previous derivation,
but with $\intdep$ and $\sfdep$ replaced by
$\intdeppert$ and $\sfdeppert$, respectively,
\phantomsection\begin{IEEEeqnarray}{rCl}
\label{eq25}
    \rf\left(\bm{r}\right)&=& \int \mathcal{B}\left[\exttot\arga\, \sfdeppert\arga\, \expo{-\opdepth\arga}\right] \,\ud\Omega\, ,
\end{IEEEeqnarray}
and the response function 
to the integrated 
line strength is 
$\rfint=\int\rf\,\ud\nu$.

Response functions can be used
to study the sensitivity of a spectral line
to specific atmospheric variables
\citep{1991A&amp;A...250..445A}.
To identify the line forming regions, however,
contribution functions must be used.

\subsection{Comparison to the plane-parallel line flux contribution function}
\label{comparison}

In the limit of plane-parallel symmetry,
the integrand in \eqn{eq14} loses 
its dependence on the azimuthal
angle $\phi$:
$\exttot\argument\rightarrow\exttot\left(z;\mu\right)$,
$\sfdep\argument\rightarrow\sfdep\left(z;\mu\right)$,
$\opdepth\argument\rightarrow\opdepthr\left(z;\mu\right)/\mu$,
where $z$ is the geometrical height
and $\opdepthr$ is the radial optical depth
The 3D contribution function $\cf$ thus
tends to a plane-parallel contribution function $\cfpp$,
\phantomsection\begin{IEEEeqnarray}{rCl}
\label{eq25}
    \cfpp\left(z\right) &=& 2\mathpi \int \exttot\left(z;\mu\right) \sfdep\left(z;\mu\right)\, \expo{-\opdepthr\left(z;\mu\right)/\mu} \, \ud \mu\, .
\end{IEEEeqnarray}
This expression is the same\footnote{After expressing the contribution function in that paper 
with respect to geometrical height
instead of radial optical depth, 
they are the same to a factor of $2\mathpi$, which arises from 
those authors integrating over spherical polar
angle $\mu$ instead
of solid angle $\Omega$}
as that 
derived by \citet{1996MNRAS.278..337A} 
in the context of 1D models,
i.e.~with the implicit
assumption of plane-parallel symmetry.

\section{Example: 3D non-LTE spectral line formation}
\label{example}

\begin{figure}
\begin{center}
\includegraphics[scale=0.31]{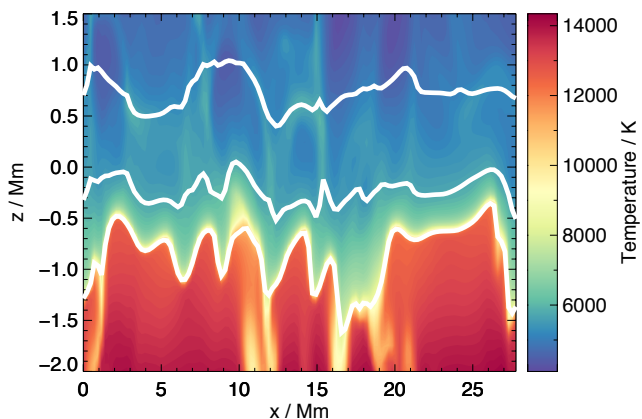}
\caption{Material temperature
in a vertical slice of a temporal snapshot
of a 3D hydrodynamic \textsc{stagger} model atmosphere
\citep{2013A&amp;A...557A..26M}.
The snapshot has effective temperature
$T_{\mathrm{eff}}\approx6430\,\mathrm{K}$, 
logarithmic surface gravity (in CGS units)
$\log_{10}g=4$, and solar-value abundances.
Contours of standard logarithmic optical depth
$\lgtfive=-3\text{,}-1\text{, and }1$ (from top to bottom)
are overdrawn.}
\label{fig2}
\end{center}
\end{figure}

\begin{figure*}
\begin{center}
\includegraphics[scale=0.31]{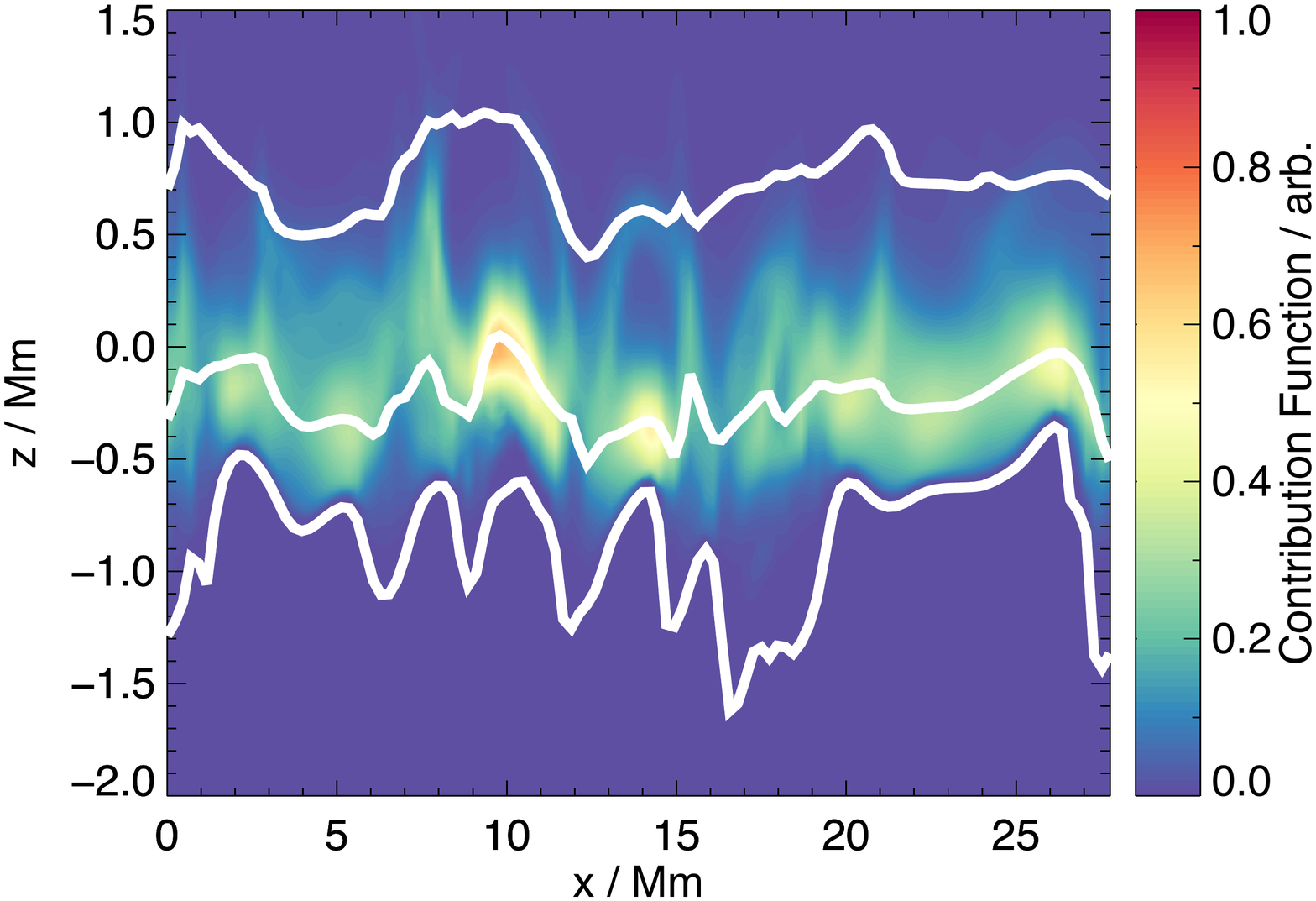}
\includegraphics[scale=0.31]{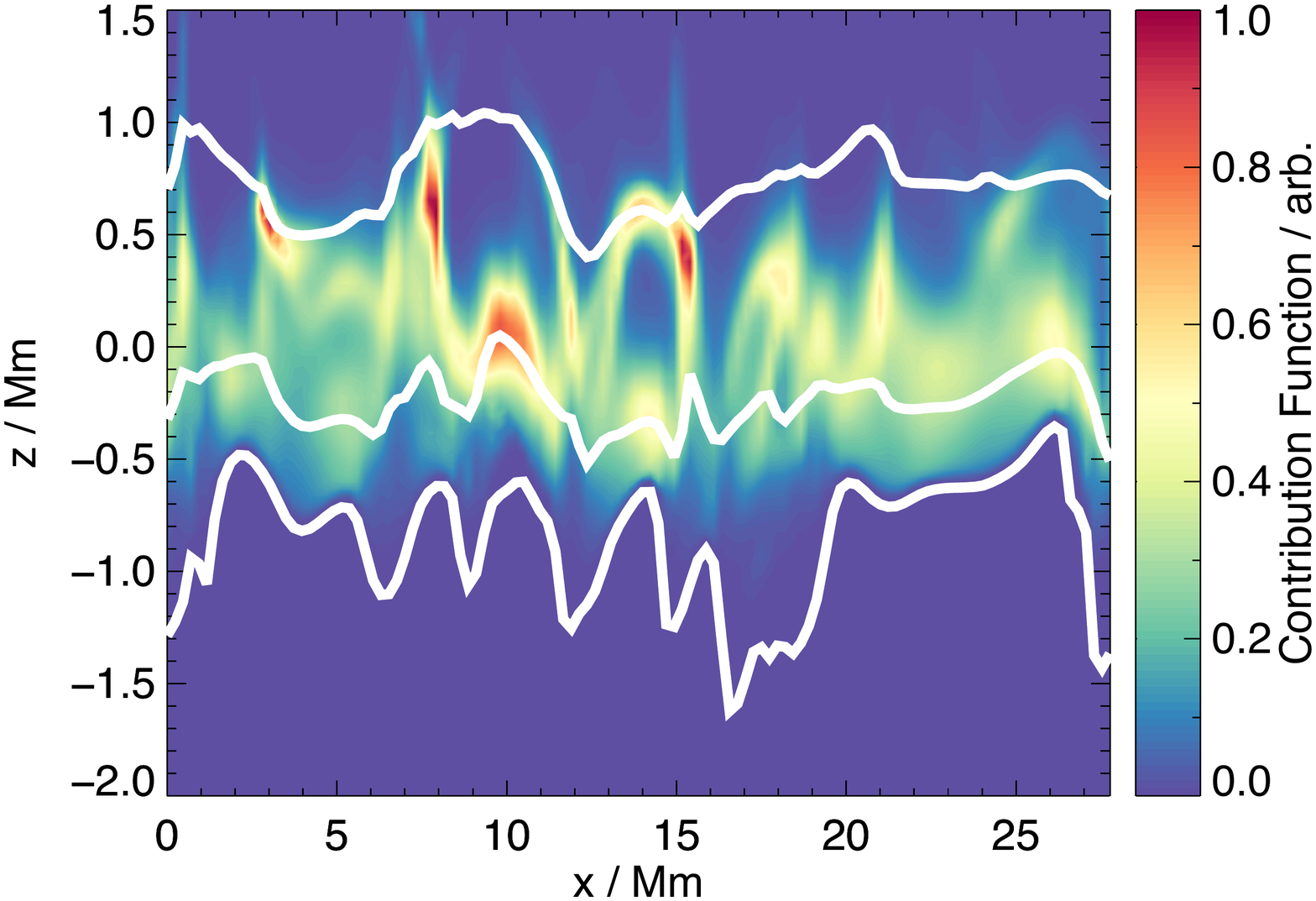}
\caption{The contribution function
$\mathcal{C_{L}^{\mathrm{box}}}$ across
the oxygen triplet
(777.25nm to 777.85nm in vacuum)
corresponding to the snapshot
slice in Fig.~\ref{fig2},
in LTE (left)
and in non-LTE (right).
These quantities are expressed
in the same arbitrary units.
Contours of standard logarithmic optical depth
$\lgtfive=-3\text{,}-1\text{, and }1$ (from top to bottom)
are overdrawn.}
\label{fig3}
\end{center}
\end{figure*}

The high excitation permitted \triplet~lines
are known to 
show departures from 
local thermodynamic equilibrium (LTE) 
\citep[LTE; ][]{1974A&amp;A....31...23S,1995A&amp;A...302..578K,2009A&amp;A...500.1221F}.
It is interesting to consider how the 
lines form within the atmosphere
when LTE is imposed,
and to see what happens
once this assumption is relaxed.

To that end, the contribution function $\cfint$
was implemented into
the 3D non-LTE radiative transfer code 
\textsc{multi3d} \citep{2009ASPC..415...87L}.
The contribution function for the \triplet~lines
was calculated 
using a model oxygen atom based on 
those used by  
\citet{1993ApJ...402..344C},
\citet{1993A&amp;A...275..269K}
and \citet{2009A&amp;A...500.1221F}.
A temporal snapshot of a 3D hydrodynamic
model atmosphere taken from the 
\textsc{stagger}-grid 
\citep{2011JPhCS.328a2003C,2013A&amp;A...557A..26M}
was used.
The model was of a typical turn-off star,
with effective temperature
$T_{\mathrm{eff}}\approx6430\,\mathrm{K}$, 
logarithmic surface gravity (in CGS units)
$\log_{10}g=4$, and solar-value abundances
\citep{2009ARA&amp;A..47..481A}.
The solid angle was sampled
using Carlson's quadrature set A4 \citep{carlson1963numerical}.

\fig{fig2} shows the temperature structure in
a vertical slice of the snapshot. 
While the absolute geometrical depth and width scales
are arbitrary, zero geometrical depth
is roughly located at the photosphere. 
Just below this depth is the top of the convection zone:
hot, light upflows, observed as wide granules,
turnover to form cool, dense downflows, observed as 
narrow intergranular lanes
\citep{2013A&amp;A...557A..26M}.
Higher up the atmosphere,
reversed granulation patterns
can be observed:
the material above the hot, light granules
expands adiabatically and cools more efficiently,
than the material above the intergranular lanes
-- a detailed discussion
can be found in the 
appendix of \citet{2013A&amp;A...560A...8M}.

The LTE and non-LTE contribution functions in this snapshot slice
are shown in \fig{fig3}.
They are both normalized such that the maximum value
of the non-LTE contribution function is 1.0.
The contribution functions
reveal that the formation of the lines is qualitatively
similar in the two cases. 
There is no contribution at large optical depths.
This can be attributed
the attenuation factor $\expo{-\opdepth}$
in the expression for the
contribution function, \eqn{eq15}:
deep within the atmosphere,
photons are more likely to be absorbed
than to penetrate the atmosphere and 
reach the observer.
Line formation is also inefficient
in the optically thin layers.
This is by virtue of the line opacity
which appears in \eqn{eq15}:
in these layers, $\extline\approx0$,
so that there is little line absorption.
Between these two extremes,
line formation becomes possible 
once the optical depth becomes
small, and the factor 
$I_{\mathrm{c}}-S_{\mathrm{l}}$ 
appearing in the effective source function
becomes non-zero
i.e.~once the re-emitted light 
is no longer equal to the absorbed light
\citep{1986A&amp;A...163..135M,1996MNRAS.278..337A}.

\fig{fig3} shows that imposing LTE inhibits the formation
of the \triplet~lines.
Photon losses in the 
triplet lines themselves
\citep{2005ARA&amp;A..43..481A}
drive departures from LTE.
The line opacity is larger,
and the line source function is smaller,
than their LTE counterparts \citep{2009A&amp;A...500.1221F}.
This leads to a significant strengthening of the lines,
correlated with the reversed granulation
patterns seen in \fig{fig2}.
The equivalent width ratio is 
$W^{\text{3D non-LTE}}/W^{\text{3D LTE}}\approx1.5$.

\section{Summary}
\label{conclusion}

Flux profiles observed from stars
have contributions from all parts of its atmosphere:
thus, the line flux contribution function
is a function of 3D space.
In this paper I have shown how to derive 
the contribution function
to the absolute flux depression
that emerges from 3D
box-in-a-star model stellar atmospheres.
The result can be used like other 1D contribution functions
\citep{1986A&amp;A...163..135M,1996MNRAS.278..337A}
to help one visualize and understand spectral line formation in 
stellar atmospheres.

\section*{Acknowledgements}
\label{acknowledgements}
I thank Martin Asplund and Remo Collet 
for advice on the original manuscript,
and Jorrit Leenaarts for providing \textsc{multi3d}.
This research was undertaken with the 
assistance of resources from the 
National Computational Infrastructure (NCI),
which is supported by the Australian Government.

\bibliographystyle{mn2e}
\bibliography{/Users/ama51/Documents/work/papers/allpapers/bibl.bib}

\label{lastpage}
\end{document}

%% file: journals.tex
\def\apjl{ApJ}
\def\apjs{ApJS}
\def\ao{Appl.~Opt.}

\def\aap{A\&A}

%% file: paper.bbl
\begin{thebibliography}{}

\bibitem[\protect\citeauthoryear{{Achmad}, {de Jager} \&
  {Nieuwenhuijzen}}{{Achmad} et~al.}{1991}]{1991A&amp;A...250..445A}
{Achmad} L.,  {de Jager} C.,    {Nieuwenhuijzen} H.,  1991, \aap, 250, 445

\bibitem[\protect\citeauthoryear{{Albrow} \& {Cottrell}}{{Albrow} \&
  {Cottrell}}{1996}]{1996MNRAS.278..337A}
{Albrow} M.~D.,  {Cottrell} P.~L.,  1996, \mnras, 278, 337

\bibitem[\protect\citeauthoryear{{Asplund}}{{Asplund}}{2005}]{2005ARA&amp;A..43..481A}
{Asplund} M.,  2005, \araa, 43, 481

\bibitem[\protect\citeauthoryear{{Asplund}, {Grevesse}, {Sauval} \&
  {Scott}}{{Asplund} et~al.}{2009}]{2009ARA&amp;A..47..481A}
{Asplund} M.,  {Grevesse} N.,  {Sauval} A.~J.,    {Scott} P.,  2009, \araa, 47,
  481

\bibitem[\protect\citeauthoryear{{Asplund}, {Nordlund}, {Trampedach} \&
  {Stein}}{{Asplund} et~al.}{1999}]{1999A&amp;A...346L..17A}
{Asplund} M.,  {Nordlund} {\AA}.,  {Trampedach} R.,    {Stein} R.~F.,  1999,
  \aap, 346, L17

\bibitem[\protect\citeauthoryear{{Beckers} \& {Milkey}}{{Beckers} \&
  {Milkey}}{1975}]{1975SoPh...43..289B}
{Beckers} J.~M.,  {Milkey} R.~W.,  1975, \solphys, 43, 289

\bibitem[\protect\citeauthoryear{Bergemann}{Bergemann}{2014}]{Bergemann:2014aa}
Bergemann M.,  2014, Analysis of Stellar Spectra with 3-D and NLTE Models.
Springer International Publishing, pp 187--205

\bibitem[\protect\citeauthoryear{{Caccin}, {Gomez}, {Marmolino} \&
  {Severino}}{{Caccin} et~al.}{1977}]{1977A&amp;A....54..227C}
{Caccin} B.,  {Gomez} M.~T.,  {Marmolino} C.,    {Severino} G.,  1977, \aap,
  54, 227

\bibitem[\protect\citeauthoryear{{Caffau}, {Ludwig}, {Steffen}, {Ayres},
  {Bonifacio}, {Cayrel}, {Freytag} \& {Plez}}{{Caffau}
  et~al.}{2008}]{2008A&amp;A...488.1031C}
{Caffau} E.,  {Ludwig} H.-G.,  {Steffen} M.,  {Ayres} T.~R.,  {Bonifacio} P.,
  {Cayrel} R.,  {Freytag} B.,    {Plez} B.,  2008, \aap, 488, 1031

\bibitem[\protect\citeauthoryear{Carlson}{Carlson}{1963}]{carlson1963numerical}
Carlson B.~G.,  1963, Methods in Computational Physics, 1, 1

\bibitem[\protect\citeauthoryear{{Carlsson} \& {Judge}}{{Carlsson} \&
  {Judge}}{1993}]{1993ApJ...402..344C}
{Carlsson} M.,  {Judge} P.~G.,  1993, \apj, 402, 344

\bibitem[\protect\citeauthoryear{{Collet}, {Asplund} \& {Trampedach}}{{Collet}
  et~al.}{2008}]{2008MmSAI..79..649C}
{Collet} R.,  {Asplund} M.,    {Trampedach} R.,  2008, \memsai, 79, 649

\bibitem[\protect\citeauthoryear{{Collet}, {Magic} \& {Asplund}}{{Collet}
  et~al.}{2011}]{2011JPhCS.328a2003C}
{Collet} R.,  {Magic} Z.,    {Asplund} M.,  2011, Journal of Physics Conference
  Series, 328, 012003

\bibitem[\protect\citeauthoryear{{de Jager}}{{de
  Jager}}{1952}]{1952hss..book.....D}
{de Jager} C.,  1952, {The hydrogen spectrum of the sun}.
Druk: Excelsiors Foto-Offset, s-Gravenhage

\bibitem[\protect\citeauthoryear{{Dravins} \& {Nordlund}}{{Dravins} \&
  {Nordlund}}{1990}]{1990A&amp;A...228..203D}
{Dravins} D.,  {Nordlund} A.,  1990, \aap, 228, 203

\bibitem[\protect\citeauthoryear{{Fabbian}, {Asplund}, {Barklem}, {Carlsson} \&
  {Kiselman}}{{Fabbian} et~al.}{2009}]{2009A&amp;A...500.1221F}
{Fabbian} D.,  {Asplund} M.,  {Barklem} P.~S.,  {Carlsson} M.,    {Kiselman}
  D.,  2009, \aap, 500, 1221

\bibitem[\protect\citeauthoryear{{Freytag}, {Steffen}, {Ludwig},
  {Wedemeyer-B{\"o}hm}, {Schaffenberger} \& {Steiner}}{{Freytag}
  et~al.}{2012}]{2012JCoPh.231..919F}
{Freytag} B.,  {Steffen} M.,  {Ludwig} H.-G.,  {Wedemeyer-B{\"o}hm} S.,
  {Schaffenberger} W.,    {Steiner} O.,  2012, Journal of Computational
  Physics, 231, 919

\bibitem[\protect\citeauthoryear{{Gray}}{{Gray}}{1992}]{1992oasp.book.....G}
{Gray} D.~F.,  1992, {The observation and analysis of stellar photospheres.}.
Cambridge Univ. Press, Cambridge

\bibitem[\protect\citeauthoryear{{Gurtovenko}, {Ratnikova} \& {de
  Jager}}{{Gurtovenko} et~al.}{1974}]{1974SoPh...37...43G}
{Gurtovenko} E.,  {Ratnikova} V.,    {de Jager} C.,  1974, \solphys, 37, 43

\bibitem[\protect\citeauthoryear{{Hubeny} \& {Mihalas}}{{Hubeny} \&
  {Mihalas}}{2014}]{2014tsa..book.....H}
{Hubeny} I.,  {Mihalas} D.,  2014, {Theory of Stellar Atmospheres}.
Princeton Univ. Press, Princeton, NJ

\bibitem[\protect\citeauthoryear{{Kiselman}}{{Kiselman}}{1993}]{1993A&amp;A...275..269K}
{Kiselman} D.,  1993, \aap, 275, 269

\bibitem[\protect\citeauthoryear{{Kiselman} \& {Nordlund}}{{Kiselman} \&
  {Nordlund}}{1995}]{1995A&amp;A...302..578K}
{Kiselman} D.,  {Nordlund} A.,  1995, \aap, 302, 578

\bibitem[\protect\citeauthoryear{{Leenaarts} \& {Carlsson}}{{Leenaarts} \&
  {Carlsson}}{2009}]{2009ASPC..415...87L}
{Leenaarts} J.,  {Carlsson} M.,  2009, in {Lites} B.,  {Cheung} M.,  {Magara}
  T.,  {Mariska} J.,   {Reeves} K.,  eds, The Second Hinode Science Meeting:
  Beyond Discovery-Toward Understanding Vol.~415 of Astronomical Society of the
  Pacific Conference Series, {MULTI3D: A Domain-Decomposed 3D Radiative
  Transfer Code}.
p.~87

\bibitem[\protect\citeauthoryear{{Ludwig}, {Caffau}, {Steffen}, {Freytag},
  {Bonifacio} \& {Ku{\v c}inskas}}{{Ludwig} et~al.}{2009}]{2009MmSAI..80..711L}
{Ludwig} H.-G.,  {Caffau} E.,  {Steffen} M.,  {Freytag} B.,  {Bonifacio} P.,
  {Ku{\v c}inskas} A.,  2009, \memsai, 80, 711

\bibitem[\protect\citeauthoryear{{Magain}}{{Magain}}{1986}]{1986A&amp;A...163..135M}
{Magain} P.,  1986, \aap, 163, 135

\bibitem[\protect\citeauthoryear{{Magic}, {Collet}, {Asplund}, {Trampedach},
  {Hayek}, {Chiavassa}, {Stein} \& {Nordlund}}{{Magic}
  et~al.}{2013a}]{2013A&amp;A...557A..26M}
{Magic} Z.,  {Collet} R.,  {Asplund} M.,  {Trampedach} R.,  {Hayek} W.,
  {Chiavassa} A.,  {Stein} R.~F.,    {Nordlund} {\AA}.,  2013a, \aap, 557, A26

\bibitem[\protect\citeauthoryear{{Magic}, {Collet}, {Hayek} \&
  {Asplund}}{{Magic} et~al.}{2013b}]{2013A&amp;A...560A...8M}
{Magic} Z.,  {Collet} R.,  {Hayek} W.,    {Asplund} M.,  2013b, \aap, 560, A8

\bibitem[\protect\citeauthoryear{{Mein}}{{Mein}}{1971}]{1971SoPh...20....3M}
{Mein} P.,  1971, \solphys, 20, 3

\bibitem[\protect\citeauthoryear{{Nordlund}}{{Nordlund}}{1980}]{1980LNP...114..213N}
{Nordlund} A.,  1980, in {Gray} D.~F.,  {Linsky} J.~L.,  eds, IAU Colloq. 51:
  Stellar Turbulence Vol.~114 of Lecture Notes in Physics, Berlin Springer
  Verlag, {Numerical simulation of granular convection - Effects on
  photospheric spectral line profiles}.
pp 213--224

\bibitem[\protect\citeauthoryear{{Nordlund}}{{Nordlund}}{1982}]{1982A&amp;A...107....1N}
{Nordlund} A.,  1982, \aap, 107, 1

\bibitem[\protect\citeauthoryear{{Sedlmayr}}{{Sedlmayr}}{1974}]{1974A&amp;A....31...23S}
{Sedlmayr} E.,  1974, \aap, 31, 23

\bibitem[\protect\citeauthoryear{{Staude}}{{Staude}}{1972}]{1972SoPh...24..255S}
{Staude} J.,  1972, \solphys, 24, 255

\bibitem[\protect\citeauthoryear{{Trampedach}, {Asplund}, {Collet}, {Nordlund}
  \& {Stein}}{{Trampedach} et~al.}{2013}]{2013ApJ...769...18T}
{Trampedach} R.,  {Asplund} M.,  {Collet} R.,  {Nordlund} {\AA}.,    {Stein}
  R.~F.,  2013, \apj, 769, 18

\end{thebibliography}
